\documentclass[draftclsnofoot,onecolumn]{IEEEtran}
\usepackage[cmex10]{amsmath}
\usepackage[psamsfonts]{amssymb}
\usepackage[dvips]{graphicx}
\usepackage{cite}
\usepackage{bm}

\newtheorem{theorem}{Theorem}
\newtheorem{lemma}{Lemma}

\newtheorem{remark}{Remark}
\newtheorem{definition}{Definition}
\newcommand{\eqtri}{\stackrel{\bigtriangleup}{=}}
\renewcommand{\IEEEQED}{\IEEEQEDopen}
\newcommand{\x}{\bm{x}}
\newcommand{\y}{\bm{y}}
\newcommand{\w}{\bm{w}}
\newcommand{\hx}{\hat{x}}

\newcommand{\hX}{\hat{X}}
\newcommand{\hY}{\hat{Y}}
\newcommand{\hbx}{\hat{\bm x}}
\newcommand{\hby}{\hat{\bm y}}
\newcommand{\bX}{\bm{X}}

\newcommand{\hcX}{\hat{\mathcal{X}}}
\newcommand{\hcY}{\hat{\mathcal{Y}}}
\newcommand{\cX}{\mathcal{X}}
\newcommand{\cY}{\mathcal{Y}}
\newcommand{\cP}{\mathcal{P}}
\newcommand{\cM}{\mathcal{M}}
\newcommand{\cU}{\mathcal{U}}
\newcommand{\cS}{\mathcal{S}}
\newcommand{\tX}{\tilde{X}}
\newcommand{\tY}{\tilde{Y}}
\newcommand{\Int}{\mathbb{N}}
\newcommand{\abs}[1]{\left\lvert#1\right\rvert}
\newcommand{\funcabs}[1]{\lVert#1\rVert}
\newcommand{\vdist}[2]{\rho\left(#1,#2\right)}
\newcommand{\dmax}{d_{\max}}

\title{Universal Coding for Lossless and Lossy Complementary Delivery Problems}
\author{Shigeaki Kuzuoka,~\IEEEmembership{Member,~IEEE,}
 Akisato~Kimura,~\IEEEmembership{Senior~Member,~IEEE,}
 and~Tomohiko~Uyematsu,~\IEEEmembership{Senior~Member,~IEEE}%
\thanks{S.~Kuzuoka is with the Department of Computer and Communication
Sciences, Wakayama University, 930 Sakaedani, Wakayama, 640-8510 Japan
(e-mail: kuzuoka@ieee.org)}%
\thanks{A.~Kimura is with NTT Communication Science Laboratories, NTT
Corporation, 3-1 Morinosato Wakamiya, Atsugi-shi, Kanagawa, 243-0198
Japan (e-mail: research@akisato.org)}%
\thanks{T.~Uyematsu is with the Department of Communications and
Integrated Systems, Tokyo Institute of Technology, 2-12-1 Ookayama,
Meguro-ku, Tokyo, 152-8550 Japan (e-mail: uyematsu@ieee.org)}%
}
\date{\today}

\begin{document}
\maketitle

\begin{abstract}
This paper deals with a coding problem called complementary delivery,
where messages from two correlated sources are jointly encoded and each
decoder reproduces one of two messages using the other message as the
side information.  Both lossless and lossy universal complementary
delivery coding schemes are investigated. In the lossless case, it is
demonstrated that a universal complementary delivery code can be
constructed by only combining two Slepian-Wolf codes.  Especially, it is
shown that a universal lossless complementary delivery code, for which
error probability is exponentially tight, can be constructed from two
linear Slepian-Wolf codes.  In the lossy case, a universal complementary
delivery coding scheme based on Wyner-Ziv codes is proposed.  While the
proposed scheme cannot attain the optimal rate-distortion trade-off in
general, the rate-loss is upper bounded by a universal constant under
some mild conditions.  The proposed schemes allows us to apply any
Slepian-Wolf and Wyner-Ziv codes to complementary delivery coding.
\end{abstract}


\section{Introduction}\label{sec:intro}
\IEEEPARstart{T}{he} 
source coding problem for correlated information sources was initiated
by Slepian and Wolf \cite{SlepianWolf73}.  They treated the case where
two information sources are encoded separately and then reproduced at
the single destination.  Subsequently, various coding problems derived
from Slepian-Wolf coding have been considered 
(e.g.~\cite{Wyner75,AhlswedeKorner75,Sgarro77,KornerMarton77}).
Corresponding lossy coding problem was studied by Wyner and Ziv
\cite{WynerZiv76}, where they investigated the lossy coding problem when
the decoder can fully observe the side information.  While the messages
are encoded separately in Slepian-Wolf and Wyner-Ziv coding problems,
the coding problems involving joint encoding processes has been also
explored (e.g.~\cite{GrayWyner74,Yamamoto81,Yamamoto96,WynerWolfWillems02}).

\begin{figure}[tb]
\centering
\includegraphics{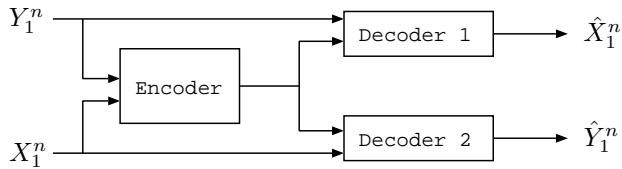}
\caption{Complementary delivery problem}\label{Fig1}
\end{figure}

This paper deals with a specific coding problem involving joint
encoding, which is called \emph{complementary delivery coding}.  The
block diagram of the complementary delivery coding is depicted in
Fig.~\ref{Fig1}.  The encoder observes messages emitted from two
correlated sources, and delivers these messages to two destinations
(decoder 1 and 2).  Each decoder reproduces one of two messages using
the other message as the side information.  Both lossless and lossy
configurations have been considered.  The lossless complementary
delivery coding can be regarded as a special case of the coding problem
investigated by Csisz{\'a}r and K{\"o}rner \cite{CsiszarKorner80} and
Wyner, Wolf and Willems \cite{WynerWolfWillems02}.  Kimura \textit{et
al.}  \cite{KimuraUyematsuKuzuoka07,KimuraUyematsuKuzuokaWatanabe}
proposed a universal coding scheme for lossless complementary delivery
based on graph coloring.  The lossy complementary delivery problem was
investigated by Kimura and Uyematsu
\cite{KimuraUyematsu-ISITA2006,KimuraUyematsu}.

In this paper, we propose universal coding schemes for lossless and
lossy complementary delivery coding problems.  At first, we propose a
simple construction of the lossless complementary delivery code
based on Slepian-Wolf codes.  The key idea of our coding scheme is as
follows.  We prepare two Slepian-Wolf codes.  One of two codes is a
Slepian-Wolf code for the source $X$ with side information $Y$, which is
used as the code from the encoder to the decoder 1.  The other code is a
Slepian-Wolf code for the source $Y$ with side information $X$, which is
used as the code from the encoder to the decoder 2.  Each source is
encoded separately by the corresponding Slepian-Wolf code, and then, the
encoder sends the summation of two codewords.  Notice that, by using the
side information $Y$, the decoder 1 can calculate the codeword from the
encoder to decoder 2.  Therefore, from the summation of two codewords,
decoder 1 can extract the codeword to reproduce $X$.  The decoder 2 can
reproduce $Y$ analogously.  The above mentioned scheme allows us to
apply any Slepian-Wolf code to lossless complementary delivery coding.
This drastically enriches the variety of complementary delivery coding.
For example, we can use universal Slepian-Wolf codes
(e.g.~\cite{Csiszar82,OohamaHan94,Uyematsu01}).  We can also apply
Slepian-Wolf codes based on the low-density parity-check matrices
(e.g.~\cite{LiverisXiongGeorghiades02,MuramatsuUyematsuWadayama05}).  In
this paper, we demonstrate that a universal lossless complementary
delivery code, for which the error probability is exponentially tight in
some rate region, can be constructed by combining linear Slepian-Wolf
codes \cite{Csiszar82}.

Next, we propose a universal lossy complementary delivery coding scheme
based on Wyner-Ziv codes \cite{WynerZiv76}.  Our scheme is universal in
the sense that it does not depend on the joint probability distribution
of the correlated sources.  While our coding scheme cannot attain the
optimal rate-distortion trade-off in general, the rate-loss is upper
bounded by a universal constant under some mild conditions.  Moreover,
our scheme allows us to construct a universal lossy complementary
delivery code by using (non-universal) Wyner-Ziv codes
(e.g.~\cite{ZamirShamaiErez02,PradhanRamchandran03,JalaliVerduWeissman07}).

\begin{figure}[tb]
\centering
\includegraphics{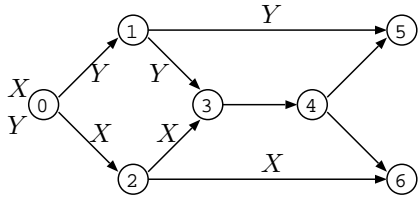}
\caption{\small
The source node $0$ observes the correlated sources $(X,Y)$, and sends
 the message to the sink nodes $5$ and $6$ over the network.
The node $3$ (resp.~$5$, $6$) corresponds to the encoder (resp.~decoder
 1,2) in Fig.~\ref{Fig1}.
}\label{Fig2}
\end{figure}

The complementary delivery coding can be regarded as a special case of
the network coding \cite{AhlswedeCaiLiYeung00,LiYeungCai03}.  Let us
consider the network depicted in Fig.~\ref{Fig2}.  The source node $0$
observes the messages emitted from the correlated sources $(X,Y)$, and
sends the message to the sink nodes $5$ and $6$ over the network.
Assume that the all edges except the edge between the nodes $3$ and $4$
have sufficiently large capacity, and thus, the output from $X$
(resp.~$Y$) can be delivered to the nodes $3$ and $6$ (resp.~$3$ and
$5$).  The problem is to find the minimum capacity between the nodes $3$
and $4$ satisfying that the codeword needed to reproduce $X$ and $Y$ can
be delivered to the nodes $5$ and $6$.  Then, this problem can be
regarded as the complementary delivery problem depicted in
Fig.~\ref{Fig1}.  The node $3$ (resp.~$5$, $6$) corresponds to the
encoder (resp.~decoder 1,2).  The coding problem of correlated sources
over a network was studied by Han \cite{Han80}.  In the recent years,
considerable attentions have been devoted to Slepian-Wolf coding over a
network
(e.g.~\cite{CristescuBLozanoVetterli05,BarrosServetto06,WuStankovicXiongKung05,RamamoorthyJainChouEffros06,HoEtal06}).
This paper shows that, for the specific network depicted in
Fig.~\ref{Fig2}, the optimal code can be constructed by only combining
two Slepian-Wolf codes.  Further, the lossy complementary delivery
investigated in this paper can be seen as a special case of lossy coding
of correlated sources over a network, which is not so studied well as
the lossless case.

This paper is organized as follows.  In Section \ref{sec:preliminary},
we introduce definitions and notations used in this paper.  In Section
\ref{sec:lossless}, lossless complementary delivery coding is
considered.  We propose a simple construction of the lossless
complementary delivery code based on Slepian-Wolf codes.  Further, we
propose another simple coding scheme which can work in a specific case.
In Section \ref{sec:lossy}, lossy complementary delivery coding is
considered.  We propose a universal lossy complementary delivery coding
scheme based on Wyner-Ziv codes.  In Section \ref{sec:conclusion}, we
present our conclusions and some directions for further work.

\section{Preliminary}\label{sec:preliminary}
We denote by $\Int$ a set of positive integers $\{1,2,\dots\}$.  For a
finite set $S$, $\abs{S}$ denotes the cardinality of $S$.  Throughout
this paper, we take all $\log$ and $\exp$ to the base 2.  We denote
random variables by upper case letters such as $X$.  Their sample values
(resp.~alphabets) are denoted by the corresponding lower case letters
such as $x$ (resp.~calligraph letters such as $\cX$).  For a random
variable $X$, $P_X$ denotes the probability distribution of $X$.
Similarly, for a pair of random variables $(X,Y)$, the joint
distribution is denoted by $P_{XY}$ and the conditional distribution of
$Y$ give $X$ is written by $P_{Y|X}$.  For each $n\in\Int$, $X^n$
denotes a random $n$-vector $(X_1,X_2,\dots,X_n)$, and
$x^n=(x_1,\dots,x_n)$ denotes a specific sample value in $\cX^n$ which
is the $n$-th Cartesian product of $\cX$.  A substring of $x^n$ is
written as $x_i^j=(x_i,x_{i+1},\dots,x_j)$ for $i\leq j$.  When the
dimension is clear from the context, vectors will be denoted by boldface
letters such as $\x\in\cX^n$.

A discrete memoryless source (DMS) is a sequence
$\bX\eqtri\{X_i\}_{i=1}^\infty$ of independent and identically
distributed (i.i.d.) copies of a random variable $X$.  For simplicity,
we call a DMS $\bX=\{X_i\}_{i=1}^\infty$ as a source $X$.  In this
paper, information theoretic quantities will be denoted following the
usual conventions of the information theory literature (see,
e.g.~\cite{Cover,CsiszarKorner}).  The \emph{entropy rate} of a source
$X$ is denoted by $H(X)$.  For a pair $(X,Y)$ of correlated sources $X$
and $Y$, the \emph{conditional entropy} of $Y$ given $X$ is denoted by
$H(Y|X)$, and the {mutual information} between $X$ and $Y$ is denoted by
$I(X;Y)$.  The \emph{relative entropy} or \emph{divergence} between two
probability distributions $P$ and $Q$ is denoted by $D(P\Vert Q)$.

For a given pair $(\x,\y)\in(\cX\times\cY)^n$ of sequences, the
\emph{joint type} of $(\x,\y)$ is defined as the empirical distribution
$Q_{\x\y}$ of $(\x,\y)$, that is,
\begin{equation*}
 Q_{\x\y}(a,b)=\frac{\abs{\{1\leq i\leq n:x_i=a,y_i=b\}}}{n}
\end{equation*}
for all $(a,b)\in\cX\times\cY$ \cite{CsiszarKorner}.  Let
$\cP_n(\cX\times\cY)$ be the set of all joint types of sequences in
$(\cX\times\cY)^n$.
By the type counting lemma \cite[Lemma 1.2.2]{CsiszarKorner}, we have
\begin{equation}
 \abs{\cP_n(\cX\times\cY)}\leq (n+1)^{\abs{\cX}\abs{\cY}}.
\label{eq:counting-lemma}
\end{equation}
Hence, we can define an injection
$\iota_n\colon\cP_n\to\{1,2,\dots,(n+1)^{\abs{\cX}\abs{\cY}}\}$.
$\iota_n(P_{\hX\hY})$ is called the index assigned to 
$P_{\hX\hY}\in\cP_n(\cX\times\cY)$.

\section{Lossless complementary delivery}\label{sec:lossless}
\subsection{Previous results}\label{subsec:3A}
In this subsection, we formulate the lossless complementary delivery
problem and show a fundamental bound of the coding rate.

\begin{definition}
 A \emph{lossless complementary delivery code} of
 block length $n$ 
 is defined by a triple of
 mappings $(f_n,\phi_n^{(1)},\phi_n^{(2)})$ where
\begin{align*}
 f_n &\colon \cX^n\times\cY^n\to\cM_n,\\
 \phi_n^{(1)} &\colon \cM_n\times\cY^n\to\cX^n,\\
 \phi_n^{(2)} &\colon \cM_n\times\cX^n\to\cY^n,
\end{align*}
where $\cM_n=\{1,2,\dots,\funcabs{f_n}\}$ and $\funcabs{f_n}<\infty$.
\end{definition}

\smallskip

\begin{definition}
 For a given pair $(X,Y)$ of correlated sources $X$ and $Y$, a rate $R$ is said to be \emph{losslessly-achievable} if
 there exists a sequence
 $\{(f_n,\phi_n^{(1)},\phi_n^{(2)})\}_{n=1}^\infty$ of codes satisfying
\begin{align*}
 \limsup_{n\to\infty}\frac{1}{n}\log\funcabs{f_n} &\leq R,\\
 \limsup_{n\to\infty}
\Pr\left\{
X^n\neq \phi_n^{(1)}\left(f_n(X^n,Y^n),Y^n\right)
\right\}&=0,\\
 \limsup_{n\to\infty}
\Pr\left\{
Y^n\neq \phi_n^{(2)}\left(f_n(X^n,Y^n),X^n\right)
\right\}&=0.
\end{align*}
\end{definition}
\medskip

As a special case of results of \cite{CsiszarKorner80} and
\cite{WynerWolfWillems02}, it can be shown that the infimum of the
losslessly-achievable rate is given by
$\max\left\{H(X|Y),H(Y|X)\right\}$.  Kimura \textit{et
al.}\cite{KimuraUyematsuKuzuoka07} proposed the universal coding scheme
based on graph coloring which can achieve any rate
$R>\max\left\{H(X|Y),H(Y|X)\right\}$.

\begin{theorem}
[Lossless coding theorem; direct part \cite{KimuraUyematsuKuzuoka07}]
\label{the:KUKdirect}
 For a given rate $R$, there exists a sequence
 $\{(f_n,\phi_n^{(1)},\phi_n^{(2)})\}_{n=1}^\infty$ of a code such that
 for any $(X,Y)$,
\[
 \limsup_{n\to\infty}\frac{1}{n}\log\funcabs{f_n}\leq R
\]
 and
\begin{align*}
\lefteqn{
\Pr\left\{
X^n\neq\phi_n^{(1)}\left(f_n(X^n,Y^n),Y^n\right)
\right\}}\\
\lefteqn{
+
\Pr\left\{
Y^n\neq\phi_n^{(2)}\left(f_n(X^n,Y^n),X^n\right)
\right\}}\\
 &\leq \exp\left\{-n\left[\min_{P_{\hX\hY}\in\overline{\cS}_n(R)} D(P_{\hX\hY}\Vert P_{XY})-\zeta_n\right]\right\}
\end{align*}
where 
\begin{align*}
\overline{\cS}_n(R)&\eqtri\biggl\{P_{\hX\hY}\in\cP_n(\cX\times\cY):\\
&\qquad \max\{H(\hX|\hY),H(\hY|\hX)\}> R\biggr\}
\end{align*}
and
\begin{align*}
 \zeta_n&\eqtri\frac{1}{n}\left\{\abs{\cX\times\cY}\log(n+1)+1\right\}\\
&\to 0\quad(n\to\infty).
\end{align*}
\end{theorem}

\medskip
On the other hand, the next theorem shows that the error exponent of the code appeared in
Theorem \ref{the:KUKdirect} is tight.

\begin{theorem}
[Lossless coding theorem; converse part \cite{KimuraUyematsuKuzuoka07}]
For any code $(f_n,\phi_n^{(1)},\phi_n^{(2)})$
 satisfying $(1/n)\log\funcabs{f_n}=R$, we have
\begin{align}
\lefteqn{
 \Pr\left\{
X^n\neq\phi_n^{(1)}\left(f_n(X^n,Y^n),Y^n\right)
\right\}}\nonumber\\
\lefteqn{
+
\Pr\left\{
Y^n\neq\phi_n^{(2)}\left(f_n(X^n,Y^n),X^n\right)
\right\}} \nonumber\\
&\geq \exp\left\{-n\left[\min_{P_{\hX\hY}\in\overline{\cS}_n(R+\zeta_n)} D(P_{\hX\hY}\Vert P_{XY})+\zeta_n\right]\right\}.\label{eq-lossless-converse}
\end{align}
\end{theorem}

\subsection{Universal coding based on Slepian-Wolf codes}
As shown in Section \ref{subsec:3A}, the coding scheme proposed in
\cite{KimuraUyematsuKuzuoka07} is universal and optimal.  However, it
requires the exponentially large coding table.  In this subsection, we
propose a simple coding scheme based on Slepian-Wolf codes.

At first, we consider Slepian-Wolf coding problem of a source $X$ with
side information $Y$. A Slepian-Wolf code of block length $n$ for a
source $X$ with side information $Y$ is defined by a pair of mappings
$(g_n^{(1)},\psi_n^{(1)})$ where
\begin{align*}
 g_n^{(1)}&\colon \cX^n\to\bar{\cM}_n,\\
 \psi_n^{(1)}&\colon \bar{\cM}_n\times\cY^n\to\cX^n,
\end{align*}
and $\bar{\cM}_n=\{1,2,\dots,\funcabs{g_n^{(1)}}\}$.  Similarly, we can
define a Slepian-Wolf code $(g_n^{(2)},\psi_n^{(2)})$ for a source $Y$
with side information $X$.
The next lemma gives a simple construction of a lossless complementary
delivery code from Slepian-Wolf codes.

\begin{lemma}
\label{lemma:CDfromSW}
 For given Slepian-Wolf codes $(g_n^{(1)},\psi_n^{(1)})$ and
 $(g_n^{(2)},\psi_n^{(2)})$, there exists a lossless complementary
 delivery code $(f_n,\phi_n^{(1)},\phi_n^{(2)})$ such that
\begin{equation*}
\funcabs{f_n}\leq\max\{\funcabs{g_n^{(1)}},\funcabs{g_n^{(2)}}\}
\end{equation*}
and
\begin{align*}
 \lefteqn{
\Pr\left\{
X^n\neq\phi_n^{(1)}\left(f_n(X^n,Y^n),Y^n\right)
\right\}
}\\
&\leq
 \Pr\left\{
X^n\neq\psi_n^{(1)}\left(g_n^{(1)}(X^n),Y^n\right)
\right\},\\
\lefteqn{\Pr\left\{
Y^n\neq\phi_n^{(2)}\left(f_n(X^n,Y^n),X^n\right)
\right\}
}\\
&\leq
\Pr\left\{
Y^n\neq\psi_n^{(2)}\left(g_n^{(2)}(Y^n),X^n\right)
\right\}.
\end{align*}
\end{lemma}

\begin{IEEEproof}
Let $M_n=\max\{\funcabs{g_n^{(1)}},\funcabs{g_n^{(2)}}\}$ and define $f_n$, $\phi_n^{(1)}$, and $\phi_n^{(2)}$ by
\begin{align*}
 f_n(\x,\y)&\eqtri g_n^{(1)}(\x)\oplus g_n^{(2)}(\y),\\
 \phi_n^{(1)}(m,\y)&\eqtri \psi_n^{(1)}\left(
m\ominus g_n^{(2)}(\y)
,\y
\right),\\
 \phi_n^{(2)}(m,\x)&\eqtri \psi_n^{(2)}\left(
m\ominus g_n^{(1)}(\x)
,\x
\right),
\end{align*}
where $\oplus$ (resp.~$\ominus$) denotes the addition
(resp.~subtraction) in modulo $M_n$ arithmetic.
The lemma follows from the construction of the code $(f_n,\phi_n^{(1)},\phi_n^{(2)})$.
\end{IEEEproof}
\medskip

Lemma \ref{lemma:CDfromSW} allows us to apply any Slepian-Wolf code to
lossless complementary delivery problem.  This drastically enriches the
variety of complementary delivery coding.  In the remaining part of this
subsection, we demonstrate that a universal code which achieves the
optimal rate $\max\{H(X|Y),H(Y|X)\}$ can be constructed by applying
universal liner Slepian-Wolf codes \cite{Csiszar82} to Lemma
\ref{lemma:CDfromSW}.

\begin{theorem}
\label{the-lossless}
Assume that $\cX=\cY$ and
$\cX$ is a Galois field.
Fix $k$ ($k\leq n$) and let $R=(k/n)\log\abs{\cX}$.
There exists a sequence $\{(f_n,\phi_n^{(1)},\phi_n^{2})\}_{n=1}^\infty$
 of lossless complementary delivery codes such that for any
 $(X,Y)$,
\begin{align*}
 \frac{1}{n}\log\funcabs{f_n}&= R
\end{align*}
and
\begin{align*}
\lefteqn{
\Pr\left\{
X^n\neq\phi_n^{(1)}\left(f_n(X^n,Y^n),Y^n\right)
\right\}}\\
&\leq \exp\left\{
-n\left(
e_r^1(R,P_{XY})-\varepsilon_n
\right)
\right\}\\
\lefteqn{
\Pr\left\{
Y^n\neq\phi_n^{(2)}\left(f_n(X^n,Y^n),X^n\right)
\right\}}\\
&\leq 
\exp\left\{
-n\left(
e_r^2(R,P_{XY})-\varepsilon_n
\right)
\right\}
\end{align*}
where
\begin{equation*}
  \varepsilon_n\eqtri\frac{2\log(n+1)}{n}\abs{\cX}^2\abs{\cY}^2
\end{equation*}
and
\begin{align*}
\lefteqn{
 e_r^1(R,P_{XY})}\nonumber\\
&\eqtri\min_{P_{\tX\tY}}\left\{D(P_{\tX\tY}\Vert
 P_{XY})+\abs{R-H(\tX|\tY)}^{+}\right\}\\
\lefteqn{
 e_r^2(R,P_{XY})}\nonumber\\
&\eqtri\min_{P_{\tX\tY}}\left\{D(P_{\tX\tY}\Vert P_{XY})+\abs{R-H(\tY|\tX)}^{+}\right\}
\end{align*}
where the minimization is over all dummy random variables $\tX$ and
$\tY$ with joint distribution $P_{\tX\tY}$, and $\abs{t}^{+}\eqtri\max(t,0)$.
\end{theorem}

\begin{IEEEproof}
Csisz{\'a}r \cite{Csiszar82} showed that there exists a linear
Slepian-Wolf code $(g_n^{(1)},\psi_n^{(1)})$ for a source $X$ with side
information $Y$ such that $g_n^{(1)}\colon\cX^n\to\cX^k$ and for any
$(X,Y)$,
\begin{align*}
\lefteqn{
 \Pr\left\{X^n\neq\psi_n^{(1)}\left(g_n^{(1)}(X^n),Y^n\right)\right\}}\nonumber\\
&\leq\exp\left\{
-n\left(
e_r^1(R,P_{XY})-\varepsilon_n
\right)
\right\}.
\end{align*}
Similarly, there exists a linear Slepian-Wolf code
$(g_n^{(2)},\psi_n^{(2)})$ for a source $Y$ with side information
 $X$ such that
$g_n^{(2)}\colon\cY^n\to\cY^k$
and for any $(X,Y)$,
\begin{align*}
\lefteqn{
 \Pr\left\{Y^n\neq\psi_n^{(2)}\left(g_n^{(2)}(Y^n),X^n\right)\right\}
}\nonumber\\
&\leq\exp\left\{
-n\left(
e_r^2(R,P_{XY})-\varepsilon_n
\right)
\right\}.
\end{align*}
By applying two codes $(g_n^{(1)},\psi_n^{(1)})$ and
 $(g_n^{(2)},\psi_n^{(2)})$ to Lemma \ref{lemma:CDfromSW}, we have the theorem.
\end{IEEEproof}
\smallskip

\begin{remark}
\label{remark:separation} As mentioned in Section \ref{sec:intro},
Theorem \ref{the-lossless} can be seen as a result of Slepian-Wolf
coding over a specific network (see Fig.~\ref{Fig2}).  In \cite[Theorem
6]{HoEtal06}, Ho \textit{et al.} also applied the linear coding approach
in \cite{Csiszar82} to Slepian-Wolf coding over a network.  However,
there are some differences between our results and the result of
\cite{HoEtal06}.  While Ho \textit{et al.} considered more general
networks than the network depicted in Fig.~\ref{Fig2}, Theorem 6 of
\cite{HoEtal06} dealt with the special case where there exists only one
receiver.  Hence, our result, Theorem \ref{the-lossless}, cannot be
derived only by applying Theorem 6 of \cite{HoEtal06} to the network
depicted in Fig.~\ref{Fig2}.  Moreover, Lemma \ref{lemma:CDfromSW}
allows us to apply not only liner but also various Slepian-Wolf codes to
network coding.  Further, the technique used in the proof of Lemma
\ref{lemma:CDfromSW} can be applied to the lossy case which was not
concerned in \cite{HoEtal06} (see Remark \ref{remark:simple-lossy}).
\end{remark}
\smallskip

\begin{remark}
In \cite{KimuraUyematsuKuzuoka07}, the coding scheme based on graph
coloring is applied to variable-rate coding for lossless complementary
delivery problem.  In the same way as the approach in
\cite{KimuraUyematsuKuzuoka07}, our coding scheme can be also modified
and applied to variable-rate coding.  For a given pair of sequences
$(\x,\y)$ to be encoded, let $\hX$ and $\hY$ be random variables such
that $P_{\hX\hY}$ is identical to the join type of $(\x,\y)$.  Let
$R=\max\{H(\hX|\hY),H(\hY|\hX)\}$ and $(f_n,\phi_n^{(1)},\phi_n^{(2)})$
be a fixed-rate lossless complementary delivery code such that
$(1/n)\log\funcabs{f_n}=R$.  If $\x=\phi_n^{(1)}(f_n(\x,\y),\y)$ and
$\y=\phi_n^{(2)}(f_n(\x,\y),\x)$, then, the encoder sends the codeword
consisting of the flag bit ``0'', the index $\iota_n(P_{\hX\hY})$ of
$P_{\hX\hY}$, and $f_n(\x,\y)$.  This codeword can be represented by
using at most $1+\abs{\cP_n(\cX\times\cY)}+R$ bits.  On the other hand,
if $\x\neq\phi_n^{(1)}(f_n(\x,\y),\y)$ or
$\y\neq\phi_n^{(2)}(f_n(\x,\y),\x)$, then, the encoder sends the
codeword consisting of the flag bit ``1'' and $(\x,\y)$, which
can be represented by using $1+\lceil n\log\abs{\cX\times\cY}\rceil$
bits.  The overflow probability of the coding rate of this scheme can be bounded in
the same way as an error probability of fixed-rate coding (see
\cite{KimuraUyematsuKuzuoka07} for more details).  Hence, it can be
 shown that, by using Slepian-Wolf codes, we can 
construct a universal variable-rate lossless complementary delivery code
for which the coding rate is smaller than or equal to
 $\max\{H(X|Y),H(Y|X)\}$ asymptotically almost surely.
\end{remark}
\medskip

Now, we investigate the tightness of the error exponent of the proposed
scheme. It is known that
\begin{align*}
 e_r^1(R,P_{XY})=\min_{H(\tX|\tY)\geq R}D(P_{\tX\tY}\Vert P_{XY})\\
 e_r^2(R,P_{XY})=\min_{H(\tY|\tX)\geq R}D(P_{\tX\tY}\Vert P_{XY})
\end{align*}
if $R\leq R_{cr}^i$ ($i=1,2$), where $R_{cr}^i=R_{cr}^i(P_{XY})$ is the largest $R$ for
which the curve $e_r^i(R,P_{XY})$ meets its supporting line of slope one
\cite{Csiszar82}.  Hence, as a corollary of Theorem \ref{the-lossless},
we have
\begin{align}
\lefteqn{\Pr\left\{
X^n\neq\phi_n^{(1)}\left(f_n(X^n,Y^n),Y^n\right)
\right\}}\nonumber\\
\lefteqn{
+
\Pr\left\{
Y^n\neq\phi_n^{(2)}\left(f_n(X^n,Y^n),X^n\right)
\right\}}\nonumber\\
&\leq 2\exp\left\{-n
\min_{P_{\tX\tY}}
D(P_{\tX\tY}\Vert P_{XY})
-\varepsilon_n
\right\}\label{eq:exponent}
\end{align}
for $R$ such that $\max\{H(X|Y),H(Y|X)\}\leq R\leq\min_{i=1,2}R_{cr}^i$.  By
comparing \eqref{eq:exponent} with \eqref{eq-lossless-converse}, it can
be seen that the error bound \eqref{eq:exponent} is exponentially tight
for $R$ such that $\max\{H(X|Y),H(Y|X)\}\leq R\leq\min_{i=1,2}R_{cr}^i$.
On the other hand, 
the error exponent bound for large rates can be improved 
in the same way as improving the error exponent of Slepian-Wolf coding.

\begin{theorem}
Assume that $\cX=\cY$ and
$\cX$ is a Galois field.
Fix $k$ ($k\leq n$) and let $R=(k/n)\log\abs{\cX}$.
There exists a sequence $\{(f_n,\phi_n^{(1)},\phi_n^{2})\}_{n=1}^\infty$
 of lossless complementary delivery codes such that for any
 $(X,Y)$,
\begin{align*}
 \frac{1}{n}\log\funcabs{f_n}&= R
\end{align*}
and
\begin{align*}
\lefteqn{
\Pr\left\{
X^n\neq\phi_n^{(1)}\left(f_n(X^n,Y^n),Y^n\right)
\right\}}\\
&\leq \exp\left\{
-n\left(
e_{x}^1(R,P_{XY})-\varepsilon_n
\right)
\right\}\\
\lefteqn{
\Pr\left\{
Y^n\neq\phi_n^{(2)}\left(f_n(X^n,Y^n),X^n\right)
\right\}}\\
&\leq 
\exp\left\{
-n\left(
e_{x}^2(R,P_{XY})-\varepsilon_n
\right)
\right\}
\end{align*}
where
\begin{align*}
\lefteqn{
 e_x^1(R,P_{XY})}\nonumber\\
&\eqtri\min_{\tX:H(\tX)\geq
 R}\left\{E_{\tX}\left[-\log\sum_{x,y}\sqrt{P_{XY}(x,y)P_{XY}(x\ominus\tX,y)}\right]+R-H(\tX)\right\}\\
\lefteqn{
 e_x^2(R,P_{XY})}\nonumber\\
&\eqtri\min_{\tY:H(\tY)\geq R}\left\{E_{\tY}\left[-\log\sum_{x,y}\sqrt{P_{XY}(x,y)P_{XY}(x,y\ominus\tY)}\right]+R-H(\tY)\right\}
\end{align*}
where $\ominus$ denotes the subtraction in the field $\cX(=\cY)$ and $E_{\tX}$ (resp.~$E_{\tY}$) denotes the expectation with respect to $P_{\tX}$ (resp.~$P_{\tY}$).
\end{theorem}

\begin{IEEEproof}
By using Slepian-Wolf codes 
which attain the
expurgated bound \cite[Theorem 3]{Csiszar82}, we
 can prove the theorem in the same way as Theorem \ref{the-lossless}.
\end{IEEEproof}

\subsection{Binary symmetric case}\label{subsec:BSS}
In this subsection, we propose another simple coding scheme for lossless
complementary delivery problem which can work in a specific case.
Let $\cX=\cY=\{0,1\}$, and consider a binary symmetric source
 with parameter $p$ ($0\leq p\leq 1/2$), that is,
\[
 P_{XY}(xy)=
\begin{cases}
 \frac{1-p}{2},&\text{ if }x=y,\\
 \frac{p}{2},&\text{ if }x\neq y.
\end{cases}
\]
In this case, a simple universal lossless code gives an optimal lossless
complementary delivery scheme.  
For a given $\x\in\cX^n$ and
$\y\in\cY^n$, let $\w\eqtri\x\oplus\y$, where $\oplus$ denotes the
addition in modulo $2$ arithmetic. Then, $\w$ can be regarded as
an output from the source $W\eqtri X\oplus Y$, which satisfies that
$P_W(0)=1-p$ and $P_W(1)=p$.
It is well known that (see e.g.~\cite{CsiszarKorner}) there exists a universal lossless code
$(\bar{f}_n,\bar{\phi}_n)$ with rate $R$ such that
\begin{equation*}
 \lim_{n\to\infty}\Pr\left\{W^n\neq \bar{\phi}_n\left(\bar{f}_n(W^n)\right)\right\}=0
\end{equation*}
provided that $R\geq h(p)$, where $h$ is the binary entropy function
defined as $h(t)\eqtri -t\log t-(1-t)\log(1-t)$.
By using $(\bar{f}_n,\bar{\phi}_n)$, we can define the code
$(f_n,\phi_n^{(1)},\phi_n^{(2)})$ as
\begin{align*}
 f_n(\x,\y)&\eqtri \bar{f}_n(\x\oplus\y),\\
 \phi_n^{(1)}(m,\y)&\eqtri \bar{\phi}_n(m)\oplus\y,\\
 \phi_n^{(2)}(m,\x)&\eqtri \bar{\phi}_n(m)\oplus\x.
\end{align*}
By the construction of the code, this simple code
$(f_n,\phi_n^{(1)},\phi_n^{(2)})$ is universal.
Further, it achieves the optimal
rate $h(p)=\max\{H(X|Y),H(Y|X)\}$ since $h(p)=H(X|Y)=H(Y|X)$.  Especially, if
$(\bar{f}_n,\bar{\phi}_n)$ is a lossless code for which the error
exponent is tight (e.g.~a code appeared in \cite{CsiszarKorner}), then,
the error exponent of the code $(f_n,\phi_n^{(1)},\phi_n^{(2)})$ based
on $(\bar{f}_n,\bar{\phi}_n)$ is also tight, that is,
$(f_n,\phi_n^{(1)},\phi_n^{(2)})$ attains the error exponent appeared in
\eqref{eq-lossless-converse}.
Furthermore, in the same way, we
can also apply lossy codes
to construct a lossy
complementary delivery code (See the proof of Theorem \ref{the:performance1}
in Appendix \ref{subsec:proof2}).

\section{Lossy complementary delivery}\label{sec:lossy}
\subsection{Previous results}
In this subsection, we formulate the lossy complementary delivery problem
and show a fundamental bound of the coding rate.
Let $\hcX$ and
$\hcY$ be reconstruction alphabets, and
$d^{(1)}\colon\cX\times\hcX\to[0,\dmax^{(1)}]$ and
$d^{(2)}\colon\cY\times\hcY\to[0,\dmax^{(2)}]$ be single-letter distortion
functions ($\dmax^{(i)}<\infty$ $(i=1,2)$).  Then, for each $n\in\Int$, the normalized distortion
$d_n^{(1)}(\x,\hbx)$ between $\x\in\cX^n$ and $\hbx\in\hcX^n$ is defined as
\[
 d_n^{(1)}(\x,\hbx)=\frac{1}{n}\sum_{i=1}^nd^{(1)}(x_i,\hx_i).
\]
For $\y\in\cY^n$ and $\hby\in\hcY^n$, $d_n^{(2)}(\y,\hby)$ is defined
similarly. Now, we define codes for lossy complementary delivery problem.

\begin{definition}
 A \emph{lossy complementary delivery code} of block length $n$
 is defined by a triple of
 mappings $(f_n,\phi_n^{(1)},\phi_n^{(2)})$ where
\begin{align*}
 f_n &\colon \cX^n\times\cY^n\to\cM_n,\\
 \phi_n^{(1)} &\colon \cM_n\times\cY^n\to\hcX^n,\\
 \phi_n^{(2)} &\colon \cM_n\times\cX^n\to\hcY^n,
\end{align*}
where $\cM_n=\{1,2,\dots,\funcabs{f_n}\}$ and $\funcabs{f_n}<\infty$.
\end{definition}
\medskip

Next, we define the achievability of rate and the optimal rate attained
by lossy complementary delivery coding.

\begin{definition}
 For a given $(X,Y)$ and a distortion pair
 $(\Delta^{(1)},\Delta^{(2)})$, a rate $R$ is said to be \emph{$(\Delta^{(1)},\Delta^{(2)})$-achievable} if
 there exists a sequence
 $\{(f_n,\phi_n^{(1)},\phi_n^{(2)})\}_{n=1}^\infty$ of codes satisfying
\begin{align*}
 \limsup_{n\to\infty}\frac{1}{n}\log\funcabs{f_n} &\leq R,\\
 \limsup_{n\to\infty}
E_{XY}\left[
d_n^{(1)}\left(
X^n,\phi_n^{(1)}\left(f_n(X^n,Y^n),Y^n\right)
\right)\right] &\leq \Delta^{(1)},\\
 \limsup_{n\to\infty}
E_{XY}\left[
d_n^{(2)}\left(
Y^n,\phi_n^{(2)}\left(f_n(X^n,Y^n),X^n\right)
\right)\right] &\leq \Delta^{(2)},
\end{align*}
where $E_{XY}$ denotes the expectation with respect to $P_{XY}$.
\end{definition}

\smallskip

\begin{definition}
For a pair of sources $(X,Y)$ and a pair of distortions
 $(\Delta^{(1)},\Delta^{(2)})$, let
\begin{equation*}
 R^*(X,Y|\Delta^{(1)},\Delta^{(2)})\eqtri\inf\left\{R: R\text{ is }(\Delta^{(1)},\Delta^{(2)})\text{-achievable}\right\}.
\end{equation*}
\end{definition}
\medskip

Kimura and Uyematsu \cite{KimuraUyematsu-ISITA2006,KimuraUyematsu}
revealed the optimal achievable rate for the lossy complementary delivery.

\begin{theorem}
[Lossy coding theorem \cite{KimuraUyematsu-ISITA2006,KimuraUyematsu}]\label{the:kimura}
 For a given $(X,Y)$,
\begin{align*}
\lefteqn{
 R^*(X,Y|\Delta^{(1)},\Delta^{(2)})}\nonumber\\
&=\min_{P_{U|XY}}\left[\max\{I(X;U|Y),I(Y;U|X)\}\right]
\end{align*}
where the minimization is over all the auxiliary random variable $U$
satisfying the following properties:
 \begin{enumerate}
  \item $P_{XYU}(x,y,u)=P_{XY}(x,y)P_{U|XY}(u|x,y)$,
  \item $U$ takes a value over an alphabet
 $\cU$ satisfying $\abs{\cU}\leq\abs{\cX\times\cY}+2$, and 
  \item there are functions $\varphi^{(1)}\colon\cU\times\cY\to\hcX$ and
	$\varphi^{(2)}\colon\cU\times\cX\to\hcY$ satisfying
\begin{align*}
 \Delta^{(1)}&\geq E_{XYU}\left[d^{(1)}(X,\varphi^{(1)}(U,Y))\right],\\
 \Delta^{(2)}&\geq E_{XYU}\left[d^{(2)}(Y,\varphi^{(2)}(U,X))\right].
\end{align*}
\end{enumerate}
\end{theorem}

\subsection{Universal coding based on Wyner-Ziv codes}\label{subsec:universal-lossy}
The coding scheme appeared in the direct part of the proof of Theorem
\ref{the:kimura} depends on  the joint distribution $P_{XY}$ of $(X,Y)$.
We propose a lossy
complementary delivery coding scheme
which does not depend on the joint distribution.

At first, we consider Wyner-Ziv coding problem of a source
$X$ with side information $Y$ under the distortion constraint
$\Delta^{(1)}$ associated with the distortion measure
$d^{(1)}\colon\cX\times\hcX\to[0,\dmax^{(1)}]$.  A Wyner-Ziv code of
block length $n$ for
a source $X$ with side information $Y$ is defined by a
pair of mappings $(g_n^{(1)},\psi_n^{(1)})$ where
\begin{align*}
 g_n^{(1)}&\colon \cX^n\to\bar{\cM}_n,\\
\psi_n^{(1)}&\colon \bar{\cM}_n\times\cY^n\to\hcX^n,
\end{align*}
and $\bar{\cM}_n=\{1,2,\dots,\funcabs{g_n^{(1)}}\}$.
Define 
$R_{WZ}(X,Y|d^{(1)},\Delta^{(1)})$ by
\begin{equation*}
 R_{WZ}(X,Y|d^{(1)},\Delta^{(1)})\eqtri\min_{P_{U|X}}\left\{I(X;U)-I(Y;U)\right\}
\end{equation*}
where the minimization is over all the
random variables $U$ satisfying the following properties:
\begin{enumerate}
 \item 
$P_{XYU}(x,y,u)=P_{U|X}(u,x)P_{XY}(x,y)$,
 \item $\abs{\cU}\leq\abs{\cX}+1$, and
 \item there exists a function $\varphi\colon\cU\times\cY\to\hcX$
       satisfying that
\begin{equation}
 E_{XYU}[d^{(1)}(X,\varphi(U,Y))]\leq \Delta^{(1)}.\label{eq:varphi}
\end{equation}
\end{enumerate}
To simplify the notation, we denote $R_{WZ}(X,Y|d^{(1)},\Delta^{(1)})$
by $R_{WZ}^{(1)}(\Delta^{(1)},P_{XY})$.  It is known that the optimal
coding rate which can be achieved by a 
Wyner-Ziv code for a source $X$ with side information $Y$ 
under the distortion constraint $\Delta^{(1)}$ is given by $R_{WZ}^{(1)}(\Delta^{(1)},P_{XY})$.

\begin{theorem}
[\cite{WynerZiv76,CsiszarKorner}]\label{the-wyner-ziv-76}
For any $\delta>0$ and any $(X,Y)$, there exists
 $l_0=l_0(\delta,\dmax^{(1)},|\cX|,|\cY|)$ such that for any $l\geq l_0$
 there exists a code $(g_l^{(1)},\psi_l^{(1)})$ satisfying
\begin{equation*}
  \frac{1}{l}\log\funcabs{g_l^{(1)}}\leq R_{WZ}^{(1)}(\Delta^{(1)},P_{XY})+\delta
\end{equation*}
and
\begin{align*}
\Pr\left\{
d_l^{(1)}\left(X^l,\psi_l^{(1)}\left(g_l^{(1)}(X^l),Y^l\right)\right)>\Delta^{(1)}
\right\}
&\leq \delta.
\end{align*}
\end{theorem}

\smallskip

\begin{remark}
While $(g_l^{(1)},\psi_l^{(1)})$ depends on $(X,Y)$, the virtue of the
 method of types \cite{CsiszarKorner} allows us to choose the block size $l$ which depends
 only on $\delta$,$\dmax^{(1)}$,$|\cX|$, and $|\cY|$.
\end{remark}

\medskip
In a similar manner to the above discussion, we can consider Wyner-Ziv
coding problem of a source $Y$ with side information $X$
under the distortion constraint
$\Delta^{(2)}$ associated with the distortion measure
$d^{(2)}\colon\cY\times\hcY\to[0,\dmax^{(2)}]$.
We denote $R_{WZ}(Y,X|d^{(2)},\Delta^{(2)})$ by 
$R_{WZ}^{(2)}(\Delta^{(2)},P_{XY})$.

Now, we describe a universal lossy complementary delivery coding scheme based
on Wyner-Ziv codes.
Fix $\gamma>0$ and $R>0$.
Choose $\delta>0$ such that $\delta<\gamma/4$ and
$4\delta\dmax^{(i)}<\gamma$ ($i=1,2$).
By Theorem \ref{the-wyner-ziv-76}, we can choose
$l=l(\delta,\dmax^{(1)},\dmax^{(2)},\abs{\cX},\abs{\cY})$ sufficiently large so that,
for any correlated sources $(\hX,\hY)$, there are $(g_l^{(1)},\psi_l^{(1)})$ and
$(g_l^{(2)},\psi_l^{(2)})$ satisfying
\begin{equation}
 \frac{1}{l}\log\funcabs{g_l^{(i)}}\leq
  R_{WZ}^{(i)}(\Delta^{(i)},P_{\hX\hY})+\delta,\quad i=1,2
\label{eq:rate-code4type}
\end{equation}
and
\begin{equation}
 \Pr\{\hX^l\hY^l\notin\Gamma_{l}(P_{\hX\hY})\}\leq 2\delta
\label{eq:dist-code4type}
\end{equation}
where
\begin{align*}
 \Gamma_{l}(P_{\hX\hY})
&\eqtri\Biggl\{(x^l,y^l):\\
&\qquad d_l^{(1)}\left(x^l,\psi_l^{(1)}(g_l^{(1)}(x^l),y^l)\right)\leq\Delta^{(1)},\\
&\qquad d_l^{(2)}\left(y^l,\psi_l^{(2)}(g_l^{(2)}(y^l),x^l)\right)\leq\Delta^{(2)}
\Biggr\}.
\end{align*}
Note that $(g_l^{(i)},\psi_l^{(i)})$ ($i=1,2$) may depend on $P_{\hX\hY}$.
Especially, for each joint type $P_{\hX\hY}$, we
can choose the pair of codes $\{(g_l^{(i)},\psi_l^{(i)})\}_{i=1,2}$
satisfying \eqref{eq:rate-code4type} and \eqref{eq:dist-code4type}.
For each $n\in\Int$, fix a correspondence between 
$\cP_{n}(\cX\times\cY)$ and
the set of pairs of Wyner-Ziv codes so that 
the pair $\{(g_l^{(i)},\psi_l^{(i)})\}_{i=1,2}$ corresponding to $P_{\hX\hY}\in\cP_{n}(\cX\times\cY)$
 satisfies
\eqref{eq:rate-code4type} and \eqref{eq:dist-code4type}
\footnote{If there are two or more pairs of codes satisfying
\eqref{eq:rate-code4type} and \eqref{eq:dist-code4type} for a joint type
$P_{\hX\hY}$, then choose one of them arbitrarily and assign it to $P_{\hX\hY}$}.
Let $\bar{M}_l\eqtri 2^{l(R+3\gamma/4)}$.
Let $n$ be so large that $n> l$ and
$(\abs{\cX}\abs{\cY}/n)\log(n+1)<\gamma/4$.
In the followings, we assume that $n=Tl$ ($T\in\Int$) for simplicity.

At first, we describe the encoding scheme.
For a given $(x^n,y^n)$, find $P_{\hX\hY}\in\cP_{n}$ such that
\footnote{If there are two or more joint types satisfying the
conditions, choose one of them arbitrarily.}
\begin{equation}
 \max_{i=1,2}R_{WZ}^{(i)}(\Delta^{(i)},P_{\hX\hY})\leq R+\gamma/2
\label{eq:encoder-Gamma1}
\end{equation}
and
\begin{equation}
\abs{
 \left\{t:
  (x_{tl+1}^{(t+1)l},y_{tl+1}^{(t+1)l})\notin\Gamma_l(P_{\hX\hY})\right\}
}
\leq
 4\delta T.
\label{eq:encoder-Gamma2}
\end{equation}
Note that $P_{\hX\hY}$ is not necessarily the joint type of $(x^n,y^n)$.
If there is no $P_{\hX\hY}\in\cP_{n}$ satisfying
\eqref{eq:encoder-Gamma1} and \eqref{eq:encoder-Gamma2},
then error is declared.
If there exists $P_{\hX\hY}\in\cP_{n}$ satisfying
\eqref{eq:encoder-Gamma1} and \eqref{eq:encoder-Gamma2}, then find the
pair of Wyner-Ziv codes corresponding to $P_{\hX\hY}$, that is, 
$\{(g_l^{(i)},\psi_l^{(i)})\}_{i=1,2}$
 satisfying
\eqref{eq:rate-code4type} and \eqref{eq:dist-code4type}.
Parse $(x^n,y^n)$ into $T$ blocks of size
$l$, and then, encode each block as
\begin{equation*}
 m_t=g_l^{(1)}(x_{tl+1}^{(t+1)l})\oplus g_l^{(2)}(y_{tl+1}^{(t+1)l}),
\quad t=0,\dots,T-1
\end{equation*}
where $\oplus$ denotes the addition in modulo $\bar{M}_l$ arithmetic
(Note that $\bar{M}_l\geq\max_{i=1,2}\funcabs{g_l^{(i)}}$).  Then, the
codeword assigned to $(x^n,y^n)$ is
$(\iota(P_{\hX\hY}),m_0,\dots,m_{T-1})$.  Since
\eqref{eq:counting-lemma} holds, the codeword can be described by using
$n(R+\gamma)$ bits because
\begin{equation*}
 \log(n+1)^{\abs{\cX}\abs{\cY}}+T\log \bar{M}_l\leq n(R+\gamma).
\end{equation*}

Next, we describe the decoding scheme.  We only describe the decoder
$\phi_n^{(1)}$ which outputs the reproduction sequence $\hx^n\in\hcX^n$
by using the codeword $(\iota(P_{\hX\hY}),m_0,\dots,m_{T-1})$ and the
side information $y^n$.  The decoder $\phi_n^{(2)}$ can be defined
analogously.
$\phi_n^{(1)}$ decodes the index $\iota(P_{\hX\hY})$ at first, and then,
computes $\hx^n$ as 
\begin{equation*}
  \hx_{tl+1}^{(t+1)l}\eqtri \psi_n^{(1)}\left(
m_t\ominus g_l^{(2)}(y_{tl+1}^{(t+1)l})
,y_{tl+1}^{(t+1)l}
\right),
\quad t=0,\dots,T-1
\end{equation*}
where $\psi_n^{(1)}$ and $g_l^{(2)}$ are the mappings corresponding to
$P_{\hX\hY}$, and $\ominus$ denotes the subtraction in modulo $\bar{M}_l$ arithmetic.

The next theorem shows the performance of the coding scheme described
above.

\begin{theorem}
\label{maintheorem}
Fix $\gamma>0$ and $R>0$.
There exists a sequence $\{(f_n,\phi_n^{(1)},\phi_n^{(2)})\}_{n=1}^\infty$ of lossy complementary delivery
 codes which satisfies the following property:
If $(X,Y)$ satisfies
\begin{equation*}
 R\geq\max_{i=1,2}R_{WZ}^{(i)}(\Delta^{(i)},P_{XY})
\end{equation*}
then, for sufficiently large $n$,
\begin{equation*}
  \frac{1}{n}\log\funcabs{f_n}\leq R+\gamma
\end{equation*}
and
\begin{align*}
\Pr\left\{
d_n^{(1)}\left(
X^n,\phi_n^{(1)}\left(f_n(X^n,Y^n),Y^n\right)
\right)> \Delta^{(1)}+\gamma
\right\} &\leq\gamma,\\
\Pr\left\{
d_n^{(2)}\left(
Y^n,\phi_n^{(2)}\left(f_n(X^n,Y^n),X^n\right)
\right)> \Delta^{(2)}+\gamma
\right\} &\leq\gamma.
\end{align*}
\end{theorem}

\smallskip

\begin{remark}
\label{remark:simple-lossy}
The proposed scheme is universal in the sense that the scheme does not
depend on the probability distribution $P_{XY}$ of $(X,Y)$.  To deal
with some technical difficulties in evaluating the performance of the
code, we adopt the coding scheme which parses the sequence into blocks
of fixed length $l$ and then encodes each block.  On the other hand, if
we know the the joint distribution $P_{XY}$, we can avoid technical
difficulties.  In fact, a (non-universal) lossy complementary delivery
code can be constructed by combining two Wyner-Ziv codes
$(g_n^{(1)},\psi_n^{(1)})$ and $(g_n^{(2)},\psi_n^{(2)})$ in the same
way as a lossless complementary delivery code is constructed by
Slepian-Wolf codes (Lemma \ref{lemma:CDfromSW}).  In the lossless case,
we can use universal Slepian-Wolf codes to construct a universal
lossless complementary delivery code.  However, as long as the authors
know, no universal Wyner-Ziv code has been proposed (While universal
Wyner-Ziv coding was recently studied in
\cite{MerhavZiv06,JalaliVerduWeissman07}, it is assumed that the
conditional distribution $P_{Y|X}$ of the source is known).
\end{remark}
\medskip

The proof of the theorem will be given in Appendix \ref{subsec:proof1}.

Our coding scheme allows us to construct a universal
lossy complementary delivery coding scheme based on (non-universal) Wyner-Ziv
codes.  
Especially, we can apply practical Wyner-Ziv codes
(e.g.~\cite{ZamirShamaiErez02,PradhanRamchandran03,JalaliVerduWeissman07})
to universal lossy complementary delivery.  
However, our scheme
cannot attain the optimal rate appeared in Theorem \ref{the:kimura} in
general. 

\begin{theorem}
\label{the:performance1}
There exists $(X,Y)$ such that
\begin{equation*}
 \max_{i=1,2}R_{WZ}^{(i)}(\Delta^{(i)},P_{XY})>R^*(X,Y|\Delta^{(1)},\Delta^{(2)}).
\end{equation*}
\end{theorem}
\smallskip
\begin{remark}
In the proof of Theorem \ref{the:performance1}, we give an example where
 a simple coding scheme appeared in Section \ref{subsec:BSS} can attain the
 optimal rate $R^*(X,Y|\Delta^{(1)},\Delta^{(2)})$. See Appendix
 \ref{subsec:proof2} for more details.
\end{remark}
\medskip

On the other hand, 
we can show that
the loss of the coding rate can be bounded by a universal
constant under some conditions.
Let $\cX=\hcX=\{1,2,\dots,M^{(1)}\}$ and $\cY=\hcY=\{1,2,\dots,M^{(2)}\}$.
Suppose that $d^{(i)}$ ($i=1,2$) is a \emph{balanced} distortion measure
\cite{Berger}, that is,
\begin{equation*}
 d^{(i)}(a,\hat{a})=\bar{d}^{(i)}(a\ominus\hat{a}),\quad\forall a,\hat{a}\in\{1,\dots,M^{(i)}\}
\end{equation*}
for some $\bar{d}^{(i)}:\{1,\dots,M^{(i)}\}\to[0,\dmax^{(i)}]$,
where $\ominus$ denotes
modulo-$M^{(i)}$ subtraction.
Let $C^{(i)}$ ($i=1,2$) be the \emph{minimax capacity} \cite{Zamir96},
defined as
\begin{equation*}
C^{(i)}(\Delta^{(i)})
=\inf_{N:E_N[\bar{d}^{(i)}(N)]\leq\Delta^{(i)}}\sup_{\substack{W:W\bot
  N,\\E_W[\bar{d}^{(i)}(W)]\leq\Delta^{(i)}}}I(W;W\oplus N)
\end{equation*}
where $N$ and $W$ are random variables on $\{1,2,\dots,M^{(i)}\}$, 
$\bot$ denotes statistical independence, and 
$\oplus$ denotes
modulo-$M^{(i)}$ addition.
The next theorem gives the bound on the rate-loss of our scheme.

\begin{theorem}
\label{the:performance2}
Suppose that $\cX=\hcX$ and $\cY=\hcY$. If $d^{(1)}$ and $d^{(2)}$ are balanced
 distortion measures, then
\begin{equation*}
\max_{i=1,2}R_{WZ}^{(i)}(\Delta^{(i)},P_{XY})
\leq R^*(X,Y|\Delta^{(1)},\Delta^{(2)})+C^{(i^*)}(\Delta^{(i^*)})
\end{equation*}
where $i^*=\arg\max_{i=1,2}R_{WZ}^{(i)}(\Delta^{(i)},P_{XY})$.
\end{theorem}

\medskip

Proofs of Theorem \ref{the:performance1} and Theorem
\ref{the:performance2} will be given in Appendix \ref{subsec:proof2}.

\section{Conclusion}\label{sec:conclusion}
We proposed a universal lossless (resp.~lossy) complementary delivery
coding scheme based on Slepian-Wolf (resp.~Wyner-Ziv) codes.  It was
demonstrated that a universal lossless complementary delivery code, for
which error probability is exponentially tight, can be constructed by
only combining two linear Slepian-Wolf codes.  On the other hand,
proposed lossy complementary delivery coding scheme cannot attain the
optimal rate generally, while it does not depend on the distribution of
the source.  The rate-loss of our lossy coding scheme was evaluated.
We also propose another simple coding scheme
which can work for a binary symmetric source.

Further work includes extensions to generalized complementary delivery
network \cite{KimuraUyematsuKuzuokaWatanabe}.  Another important work is
to construct a universal lossy complementary delivery code which attains
the optimal rate.


\appendix[Proofs of Theorems]

\subsection{Proof of Theorem \ref{maintheorem}}\label{subsec:proof1}

Before proving Theorem \ref{maintheorem}, we introduce some lemmas.
In this appendix, the variational distance between $P_{XY}$ and $P_{\hX\hY}$ is denoted by
$\vdist{P_{XY}}{P_{\hat{X}\hat{Y}}}$.

\begin{lemma}
\label{lemma:Extension}
For any $\delta>0$ 
and $l\in\Int$, there exists $\epsilon_1=\epsilon_1(l,\delta,\abs{\cX},\abs{\cY})>0$ such that
 if 
\begin{equation*}
 \vdist{P_{XY}}{P_{\hat{X}\hat{Y}}}\leq\epsilon_1
\end{equation*}
then,
\begin{equation*}
 \vdist{P_{X^lY^l}}{P_{\hat{X}^l\hat{Y}^l}}\leq\delta.
\end{equation*}
\end{lemma}

\begin{IEEEproof}
If 
$\vdist{P_{XY}}{P_{\hat{X}\hat{Y}}}\leq\epsilon_1$, then for any $(x^l,y^l)\in(\cX\times\cY)^l$,
 \begin{align*}
  P_{X^lY^l}(x^l,y^l)&=\prod_{i=1}^lP_{XY}(x_i,y_i)\\
&\leq\prod_{i=1}^l\left\{P_{\hX\hY}(x_i,y_i)+\epsilon_1\right\}\\
&\leq\prod_{i=1}^lP_{\hX\hY}(x_i,y_i)+\delta(\epsilon_1,l)\\
&= P_{\hX^l\hY^l}(x^l,y^l)+\delta(\epsilon_1,l)
 \end{align*}
where $\delta(\epsilon_1,l)\to 0$ as $\epsilon_1\to 0$.
Similarly, we have $P_{X^lY^l}(x^l,y^l)\geq
 P_{\hX^l\hY^l}(x^l,y^l)-\delta(\epsilon_1,l)$.
Hence, we have the lemma.
\end{IEEEproof}

\medskip

\begin{lemma}
\label{lemma:RwzContinuous2}

\begin{enumerate}
 \item 
 For any $\gamma>0$ and any $P_{\hX\hY}$, there exists $\zeta>0$
 satisfying
\begin{equation*}
 R_{WZ}^{(i)}(\Delta^{(i)},P_{\hX\hY})\leq
  R_{WZ}^{(i)}(\Delta^{(i)}+\zeta,P_{\hX\hY})+\gamma/4,\quad i=1,2.
\end{equation*}
 \item 
 For any $\gamma>0$, $\zeta>0$ and any $(X,Y)$, there exists $\epsilon_2>0$ such
 that if $\vdist{P_{XY}}{P_{\hX\hY}}<\epsilon_2$ then
\begin{equation*}
 R_{WZ}^{(i)}(\Delta^{(i)}+\zeta,P_{\hX\hY})\leq
  R_{WZ}^{(i)}(\Delta^{(i)},P_{XY})+\gamma/4,\quad i=1,2.
\end{equation*}
\end{enumerate}
\end{lemma}

\begin{IEEEproof}
 The first part of the lemma follows from the fact that
 $R_{WZ}^{(i)}(\Delta^{(i)},P_{\hX\hY})$ is continuous in $\Delta^{(i)}$ \cite{WynerZiv76}.

By the definition of $R_{WZ}^{(1)}(\Delta^{(1)},P_{XY})$, we can choose
 $P_{U|X}$ and $\varphi$ satisfying 
\begin{equation*}
 R_{WZ}^{(1)}(\Delta^{(1)},P_{XY})=I(X;U)-I(Y;U)
\end{equation*}
and \eqref{eq:varphi}.
If $\vdist{P_{XY}}{P_{\hX\hY}}$ is sufficiently small, then 
$P_{\hX\hY}$ satisfies that
\begin{equation*}
  I(\hX;U)-I(\hY;U)\leq  I(X;U)-I(Y;U)+\gamma/4
\end{equation*}
and
\begin{equation*}
  E_{\hX\hY U}[d^{(1)}(\hX,\varphi(U,\hY))]\leq \Delta^{(1)}+\zeta.
\end{equation*}
Hence, we have
\begin{equation*}
  R_{WZ}^{(1)}(\Delta^{(1)}+\zeta,P_{\hX\hY})\leq
  R_{WZ}^{(1)}(\Delta^{(1)},P_{XY})+\gamma/4.
\end{equation*}
Similarly, we can prove that
\begin{equation*}
  R_{WZ}^{(2)}(\Delta^{(2)}+\zeta,P_{\hX\hY})\leq
  R_{WZ}^{(2)}(\Delta^{(2)},P_{XY})+\gamma/4
\end{equation*}
provided that $\vdist{P_{XY}}{P_{\hX\hY}}$ is sufficiently small.
\end{IEEEproof}

\medskip

\begin{IEEEproof}
[Proof of Theorem \ref{maintheorem}]
We prove the theorem by showing that the code defined in 
Section \ref{subsec:universal-lossy} satisfies the property appeared in the
 theorem.

Let $\delta>0$ and $l\in\Int$ be numbers satisfying the
 conditions appeared in the description of coding scheme.
Let $n=Tl\in\Int$ be sufficiently large.

Suppose that there exists $P_{\hX\hY}\in\cP_{n}$ satisfying
\eqref{eq:encoder-Gamma2}.
Then,
\begin{align*}
d_n^{(1)}\left(x^n,\phi_n^{(1)}(f_n(x^n,y^n),y^n)
\right)
&\leq
\frac{1}{n}\left\{
4T\delta l\dmax^{(1)}+Tl\Delta^{(1)}\right\}\\
&=\Delta^{(1)}+4\delta\dmax^{(1)}\\
&\leq\Delta^{(1)}+\gamma.
\end{align*}
Similarly,
\begin{equation*}
 d_n^{(2)}\left(y^n,\phi_n^{(2)}(f_n(x^n,y^n),x^n)
\right)\leq\Delta^{(2)}+\gamma.
\end{equation*}
Hence, to prove the theorem, it is sufficient to show that,
for sufficiently large $n$, we can find $P_{\hX\hY}\in\cP_{n}$ satisfying 
\eqref{eq:encoder-Gamma1} and \eqref{eq:encoder-Gamma2} with probability
 greater than $1-\gamma$.

Choose $\epsilon_1$ (resp.~$\zeta$, $\epsilon_2$)
satisfying 
Lemma  \ref{lemma:Extension} (resp.~Lemma \ref{lemma:RwzContinuous2}).
Fix $\epsilon>0$ such that $\epsilon<\epsilon_i$ ($i=1,2$).
If $n$ is sufficiently large,
there exists a joint type $P_{\hX\hY}\in\cP_{n}(\cX\times\cY)$ satisfying
\begin{equation}
 \vdist{P_{XY}}{P_{\hX\hY}}\leq\epsilon.\label{eq:distance}
\end{equation}
By Lemma
 \ref{lemma:RwzContinuous2}, we have
\begin{equation*}
 R_{WZ}^{(i)}(\Delta^{(i)},P_{\hX\hY})\leq
  R_{WZ}^{(i)}(\Delta^{(i)},P_{XY})+\gamma/2,\quad i=1,2.
\end{equation*}
Since $R\geq\max_{i=1,2}R_{WZ}^{(i)}(\Delta^{(i)},P_{XY})$, $P_{\hX\hY}$
 satisfies \eqref{eq:encoder-Gamma1}.
In the followings, we prove that $P_{\hX\hY}$ also satisfies
 \eqref{eq:encoder-Gamma2} with probability greater than $1-\gamma$.
By Lemma \ref{lemma:Extension}, \eqref{eq:dist-code4type}, and \eqref{eq:distance},
\begin{equation}
 P_{X^lY^l}(\Gamma_{l}^\complement)\leq
  P_{\hX^l\hY^l}(\Gamma_{l}^\complement)+\delta\leq 3\delta
\label{eq1:proof-maintheorem}
\end{equation}
where $\Gamma_{l}^\complement$ denotes the complement of $\Gamma_{l}(P_{\hX\hY})$.
Let $B_t$ ($t=0,1,\dots,T-1$) be random variables defined by
\begin{equation*}
 B_t\eqtri
\begin{cases}
1, & \left(X_{tl+1}^{(t+1)l},Y_{tl+1}^{(t+1)l}\right)\in\Gamma_{l}^\complement,\\
0, & \text{otherwise}.
\end{cases}
\end{equation*}
Since \eqref{eq1:proof-maintheorem} implies
$E_{X^lY^l}[B_i]\leq 3\delta$, by the law of large numbers, we have
\begin{equation*}
 \lim_{T\to\infty}\Pr\left\{\frac{1}{T}\sum_{t=0}^{T-1}B_t>3\delta+\delta\right\}=
  0.
\end{equation*}
Hence, if $n=Tl$ is sufficiently large, then we can find
$P_{\hX\hY}\in\cP_{n}$ satisfying \eqref{eq:encoder-Gamma1} and
\eqref{eq:encoder-Gamma2} with probability greater than $1-\gamma$.
\end{IEEEproof}

\subsection{Proofs of Theorem \ref{the:performance1} and Theorem \ref{the:performance2}}\label{subsec:proof2}
Before proving Theorem \ref{the:performance1} and Theorem
\ref{the:performance2}, we introduce some notations.
Let $R_{both}^{(1)}(\Delta^{(1)},P_{XY})$ be the optimal rate of the lossy source coding problem with side information at the
 encoder and the decoder \cite{WynerZiv76}, that is,
\begin{equation*}
 R_{both}^{(1)}(\Delta^{(1)},P_{XY})\eqtri\min_{P_{Z^{(1)}|XY}} I(X;Z^{(1)}|Y)
\end{equation*}
where the minimization is with respect to all random variables $Z^{(1)}$
 such that $P_{XYZ^{(1)}}(x,y,z)=P_{XY}(x,y)P_{Z^{(1)}|XY}(z|x,y)$ is
 the probability distribution on $\cX\times\cY\times\hcX$ satisfying
 $\sum_{x,y,z}d^{(1)}(x,z)P_{XYZ^{(1)}}(x,y,z)\leq\Delta^{(1)}$.
 Similarly, let
\begin{equation*}
 R_{both}^{(2)}(\Delta^{(2)},P_{XY})\eqtri\min_{P_{Z^{(2)}|XY}} I(Y;Z^{(2)}|X).
\end{equation*}
Note that
\begin{align}
\lefteqn{
 \max_{i=1,2}
R_{both}^{(i)}(\Delta^{(i)},P_{XY})}\nonumber\\
&\leq 
\min_{P_{U|XY}}\left[\max\{I(X;U|Y),I(Y;U|X)\}\right]\nonumber\\
&= R^*(X,Y|\Delta^{(1)},\Delta^{(2)})\label{eq:bound-both}
\end{align}
where $\min_{P_{U|XY}}$ is taken over $U$ satisfying the properties
appeared in Theorem \ref{the:kimura}.

\begin{IEEEproof}
[Proof of Theorem \ref{the:performance1}]
Let $\cX=\cY=\hcX=\hcY=\{0,1\}$, and consider a binary symmetric source
 with parameter $p$ ($0< p<1/2$).
Let $d^{(1)}$ and $d^{(2)}$ be the Hamming distortion measure, that is
$d^{(i)}(x,\hat{x})=0$ if $x=\hat{x}$ and $d^{(i)}(x,\hat{x})=1$ otherwise.
Let $\Delta^{(1)}=\Delta^{(2)}=\Delta$ ($\Delta< p$).

For a given $\x\in\cX^n$ and
$\y\in\cY^n$, let $\w\eqtri\x\oplus\y$, where $\oplus$ denotes the
addition in modulo $2$ arithmetic. Then, $\w$ can be regarded as
an output from the source $W\eqtri X\oplus Y$, which satisfies that
$P_W(0)=1-p$ and $P_W(1)=p$.

It is known
that \cite{Berger} there exists a lossy code $(\bar{g}_n,\bar{\psi}_n)$
with rate $h(p)-h(\Delta)$ satisfying that
\[
 \lim_{n\to\infty}\Pr\left\{d_n^{(1)}\left(W^n,\bar{\psi}_n(\bar{g}_n(W^n))>\Delta\right)\right\}=0.
\]
Based on $(\bar{g}_n,\bar{\psi}_n)$, define the code
$(f_n,\phi_n^{(1)},\phi_n^{(2)})$ as
\begin{align*}
 f_n(\x,\y)&\eqtri \bar{g}_n(\x\oplus\y),\\
 \phi_n^{(1)}(m,\y)&\eqtri \bar{\psi}_n(m)\oplus\y,\\
 \phi_n^{(2)}(m,\x)&\eqtri \bar{\psi}_n(m)\oplus\x.
\end{align*}
Since 
\begin{equation*}
 d_n^{(1)}(\x,\phi_n^{(1)}(m,\y))=d_n^{(1)}(\w,\bar{\psi}_n(\bar{g}_n(\w)))
\end{equation*}
and
\begin{equation*}
 d_n^{(2)}(\y,\phi_n^{(2)}(m,\x))=d_n^{(2)}(\w,\bar{\psi}_n(\bar{g}_n(\w))),
\end{equation*}
the code $(f_n,\phi_n^{(1)},\phi_n^{(2)})$ satisfies the distortion
constraints.
This fact indicates that
\begin{equation}
 h(p)-h(\Delta)\geq R^*(X,Y|\Delta^{(1)},\Delta^{(2)}).
\label{eq-example-3}
\end{equation}
Further, by the result of lossy source coding with side information at
 the encoder and decoder \cite{WynerZiv76}, we have
\begin{equation*}
 R_{both}^{(i)}(\Delta^{(i)},P_{XY})=h(p)-h(\Delta),\quad i=1,2.
\end{equation*}
Since \eqref{eq:bound-both} holds, we have
\begin{equation}
h(p)-h(\Delta)\leq
R^*(X,Y|\Delta^{(1)},\Delta^{(2)}).
 \label{eq-example-4}
\end{equation}
By combining \eqref{eq-example-3} and \eqref{eq-example-4}, we have
\begin{equation}
h(p)-h(\Delta)=
R^*(X,Y|\Delta^{(1)},\Delta^{(2)}).
 \label{eq-example-2}
\end{equation}
On the other hand, by the result of Wyner-Ziv coding problem
\cite{WynerZiv76}, we have for each $i=1,2$,
\begin{equation}
 R_{WZ}^{(i)}(\Delta^{(i)},P_{XY})
=\inf_{\theta,\beta}\left\{\theta[h(p*\beta)-h(\beta)]\right\}
\label{eq-example-1}
\end{equation}
where the infimum is with respect to all $\theta,\beta$ such that 
$0\leq\theta\leq 1$, $0\leq \beta< p$, and
$\Delta=\theta\beta+(1-\theta)p$.
\eqref{eq-example-1} and \eqref{eq-example-2} indicate that
$ \max_{i=1,2}R_{WZ}^{(i)}(\Delta^{(i)},P_{XY})>R^*(X,Y|\Delta^{(1)},\Delta^{(2)})$ for all $0<p<1/2$.
\end{IEEEproof}

\medskip

\begin{IEEEproof}
[Proof of Theorem \ref{the:performance2}]
Since \eqref{eq:bound-both} holds,
we can bound the rate loss as
\begin{align*}
\lefteqn{\max_{i=1,2}R_{WZ}^{(i)}(\Delta^{(i)},P_{XY})
 -R^*(X,Y|\Delta^{(1)},\Delta^{(2)})}\\
&\leq R_{WZ}^{(i^*)}(\Delta^{(i^*)},P_{XY})- 
 \max_{i=1,2}
R_{both}^{(i)}(\Delta^{(i)},P_{XY})\\
&\leq R_{WZ}^{(i^*)}(\Delta^{(i^*)},P_{XY})-R_{both}^{(i^*)}(\Delta^{(i^*)},P_{XY}).
\end{align*}
It is known that the difference $R_{WZ}^{(i^*)}(\Delta^{(i^*)},P_{XY})-R_{both}^{(i^*)}(\Delta^{(i^*)},P_{XY})$
is bounded by a universal constant $C^{(i^*)}(\Delta^{(i^*)})$ \cite{Zamir96}.
\end{IEEEproof}



\end{document}